\documentclass[9pt,twocolumn,twoside]{opticajnl}
\journal{opticajournal} 

\setboolean{shortarticle}{true}

\usepackage{lineno}
\usepackage{soul}

\title{Asynchronous–spectral fusion fluorescence microscopy for microsecond-scale behavioral dynamics}

\author[1]{Richard G. Baird}
\author[2]{Boyden Myers}
\author[2]{Erik M. Jorgensen}
\author[1,*]{Rajesh Menon}

\affil[1]{Department of Electrical and Computer Engineering, University of Utah, Salt Lake City, UT 84112, USA}
\affil[2]{Howard Hughes Medical Institute, School of Biological Sciences, University of Utah, Salt Lake City, UT 84112, USA}

\affil[*]{rmenon@eng.utah.edu}

\begin{abstract}
Event-based image sensors provide microsecond temporal resolution but lack spectral discrimination, whereas diffractive spectral imagers encode wavelength information at conventional frame rates. We introduce a fluorescence microscopy architecture that fuses asynchronous event streams with diffraction-encoded CMOS measurements to decouple temporal and spectral sampling. The system achieves $\sim$3.9~$\mu$m spatial resolution over a 0.5~mm field of view, effective temporal resolution down to 100~$\mu$s, and differentiates fluorophores whose emission peaks are separated by only 23~nm. By synchronizing and computationally merging both sensing modalities, we enable spectrally resolved tracking at 100,000~frames/s without scanning or filter switching. Applied to live \emph{C.~elegans}, the platform resolves blue-light–evoked omega turns and reveals a significant genotype-dependent difference in response latency (0.18~s in wild type vs 0.88~s in \emph{unc-101}). These results establish asynchronous–spectral fusion as a general strategy for ultrafast, multiplexed fluorescence microscopy capable of quantifying rapid behavioral dynamics.
\end{abstract}

\setboolean{displaycopyright}{false} 

\begin{document}

\maketitle

\section*{Introduction}

Fluorescence microscopy enables visualization of cellular and molecular processes, yet simultaneous high spatial, spectral, and temporal resolution remains challenging. Increasing temporal resolution reduces photon integration time and signal-to-noise ratio, while spectral discrimination typically requires sequential filtering or scanning. Overcoming these coupled constraints is essential for studying rapid biological dynamics.

We previously demonstrated diffraction-encoded spectral fluorescence imaging capable of resolving closely overlapping emission spectra at ultrahigh temporal rates using fluorescent beads \cite{baird_dynamic_2025}. However, extending this capability to living specimens introduces reduced photon counts, non-rigid motion, and the need for robust spatio-temporal alignment.

Dynamic-vision sensors (DVS) also referred to as event-based image sensors record asynchronous brightness changes with microsecond latency, enabling temporal bandwidth far beyond frame-based cameras. However, DVS lack intrinsic spectral discrimination. Conversely, diffraction-based spectral imaging provides wavelength separation but operates at conventional frame rates. These complementary characteristics motivate a sensing architecture in which spectral encoding and temporal sampling are decoupled and subsequently fused computationally.

We apply this architecture to \emph{C.~elegans}, a model organism for neural and behavioral dynamics \cite{obernosterer_post-transcriptional_2006,gao_aging_2024,leung_caenorhabditis_2008}. Blue light is known to elicit avoidance behaviors, including canonical omega turns, mediated by defined neural circuits \cite{hori2018off,dunkel2025neurons}. Resolving the onset and kinetics of these responses requires temporal precision beyond conventional frame-based fluorescence imaging.

Using step-modulated blue illumination, we probe stimulus-evoked behavior while simultaneously resolving fluorophores, whose emission peaks are separated by only 23~nm. The fused architecture enables quantification of omega-turn latency with sub-millisecond temporal granularity, revealing a significant delay between wild-type and \emph{unc-101} animals. These experiments demonstrate that asynchronous–spectral sensor fusion enables unified spatial, spectral, and temporal imaging in dynamic biological systems while providing access to behavioral kinetics inaccessible to conventional approaches.

\section*{Operating Principle}

\begin{figure}[htb!]
\centering
\fbox{\includegraphics[width=\linewidth]{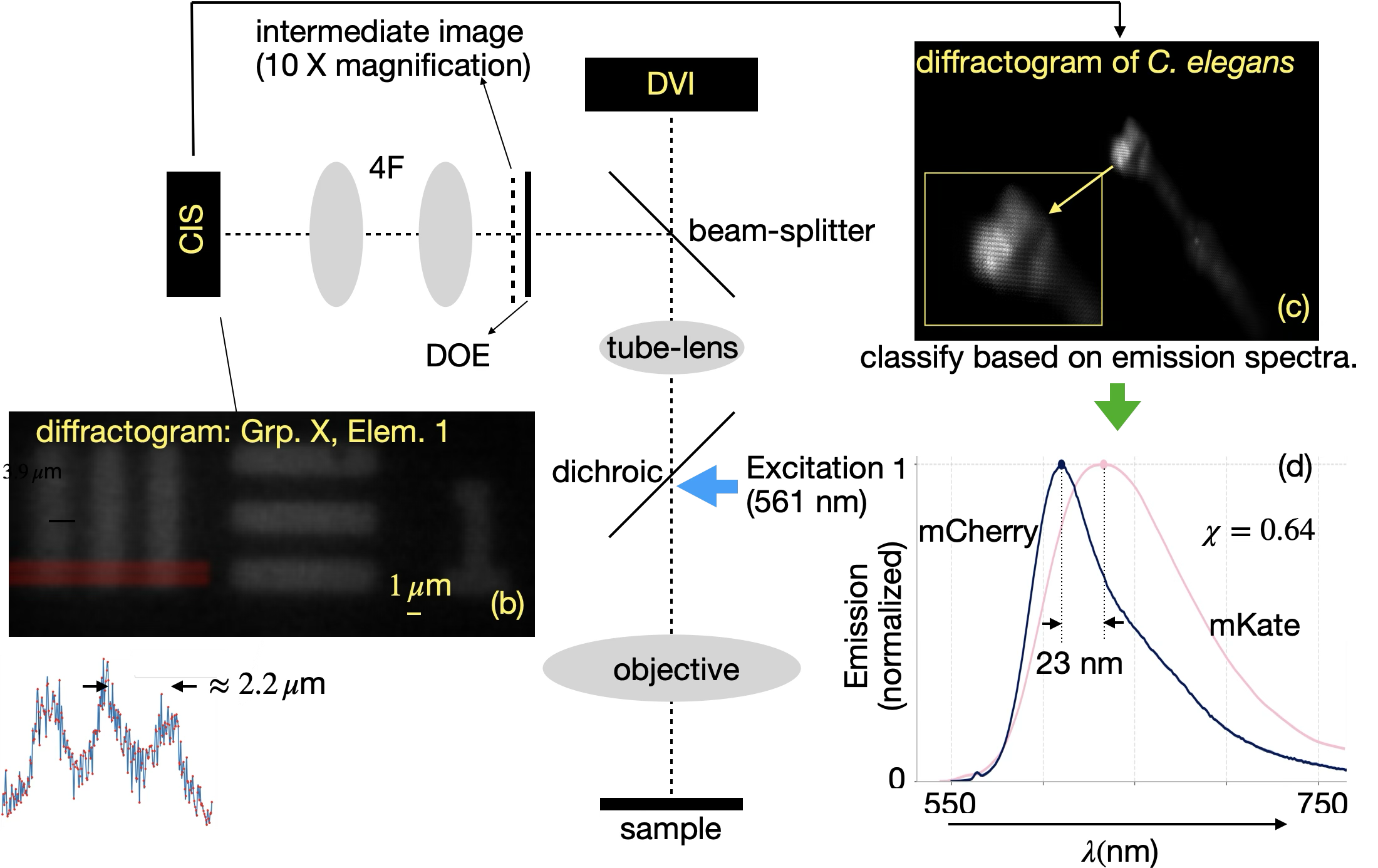}}
\caption{\textbf{Sensor-fusion fluorescence microscope.}
(a) A 50/50 beam splitter directs fluorescence simultaneously to a CMOS image sensor (CIS) and a dynamic vision sensor (DVS). A diffractive optical element (DOE) in the CIS arm encodes wavelength-dependent structure into a diffractogram relayed by a 1:1 4F system, while the DVS records asynchronous events from the same field of view.  
(b) Diffractogram of a USAF target demonstrating $\sim$2.2~$\mu$m diffraction-limited resolution.  
(c) A neural network performs pixel-level classification between spectrally overlapping fluorophores (mKate2 and mCherry; emission-peak separation = 23~nm, $>$62\% overlap).}
\label{fig:setup}
\end{figure}

The architecture builds on diffraction-encoded spectral imaging \cite{baird_dynamic_2025} and extends it to living specimens through asynchronous–spectral fusion. The microscope (Fig.~\ref{fig:setup}) employs a 10$\times$ infinity-corrected objective (NA = 0.25) with a $496.8~\mu$m field of view and diffraction-limited resolution of $1.34~\mu$m. A 50/50 beam splitter directs emission to two detection paths.

In the CMOS path (Thorlabs CS165MU), fluorescence passes through a diffractive optical element (DOE), producing wavelength-dependent diffractograms relayed by a 1:1 4F system. This approach follows prior work in computational spectroscopy and multispectral imaging \cite{majumder2025high,wang2017computational,wang2015ultra,wang2014computational,wang2014computational2} and preserves photon efficiency.

In the second path, a dynamic vision sensor (DVS; Prophesee PPS3MVCD) records asynchronous intensity changes with microsecond latency. The DVS provides high-temporal-bandwidth motion tracking but no spectral discrimination.

Diffractograms are processed using a neural network for pixel-level fluorophore classification. Spectral labels are temporally synchronized and assigned to DVS trajectories, yielding spectrally resolved dynamics at event-camera timescales. Excitation was provided by a 561~nm CW laser; a 467~nm LED was used for stimulation experiments as described below. Additional details are provided in the Supplement.

\section*{Results}

mKate2 and mCherry were selected for their photostability, far-red emission, and compatibility with \emph{C.~elegans} imaging \cite{shen_red_2015, shcherbo2007bright, shaner2004improved, shaner2008advances}. Both mature efficiently and remain fluorescent under physiological conditions \cite{yemini2021neuro, stirman2011real}. Their emission peaks differ by 23~nm with substantial spectral overlap, providing a stringent test of spectral discrimination \cite{wang2015ultra, majumder2025high}.

Due to their relatively low brightness \cite{shaner_improved_2004, shcherbo_bright_2007}, a 100~ms exposure was used on the CIS to obtain high-SNR diffractograms. The DVS, with $>$120~dB dynamic range, records asynchronous intensity changes rather than integrated frames, enabling microsecond-resolved tracking. Spectral identity determined at the CIS frame rate was fused with DVS trajectories to produce unified spatio-temporal reconstructions.

Calibration was performed using a 1951 USAF resolution target (Fig.~\ref{fig:setup}b) to verify diffraction-limited performance and measure spatial alignment between sensors \cite{baird_dynamic_2025}. Static specimens were used for training and validation of spectral classification.

\begin{figure}[htb!]
\centering
\fbox{\includegraphics[width=\linewidth]{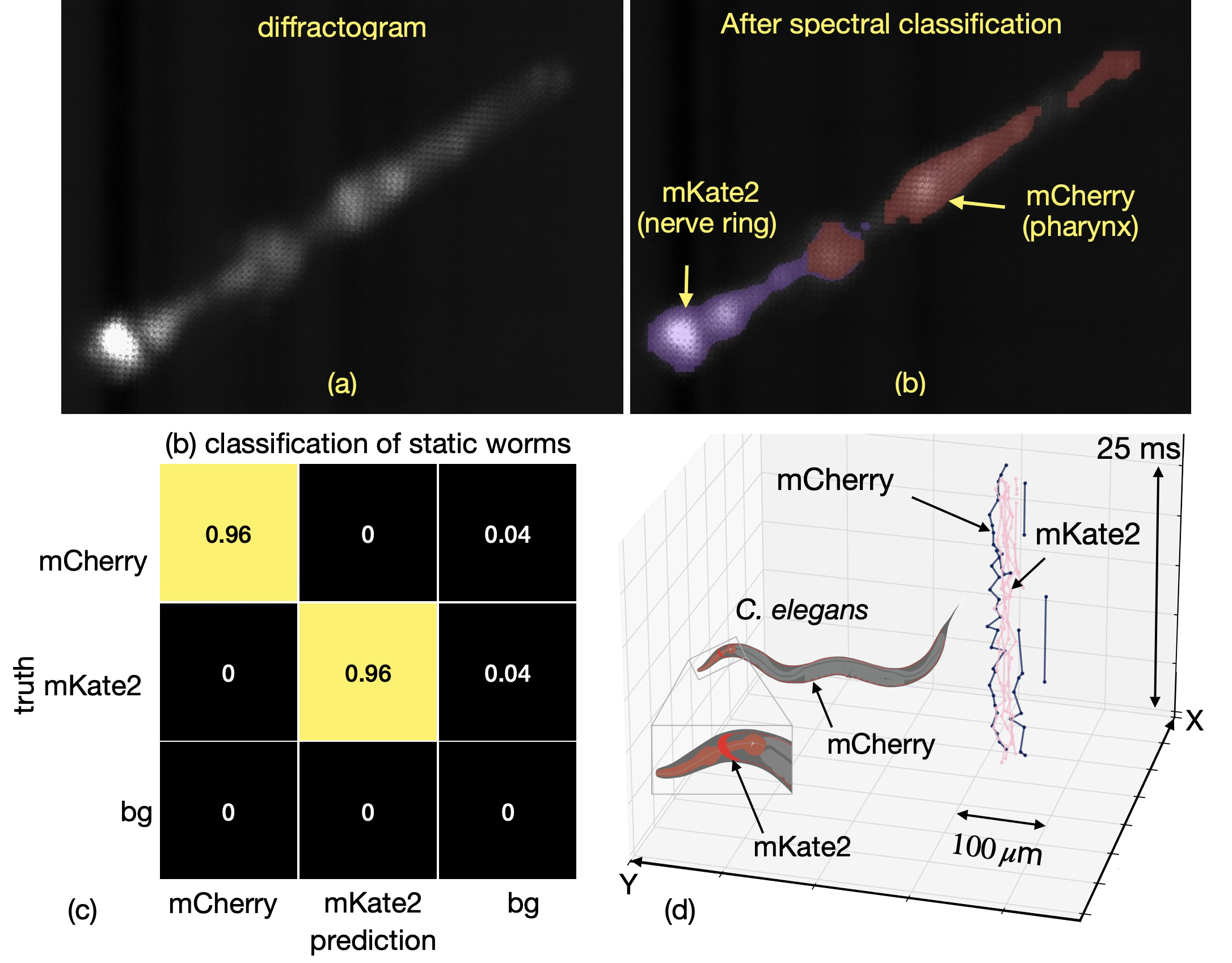}}
\caption{\textbf{Spectral classification and ultra-fast tracking.}
(a) Representative diffractogram from a static specimen. 
(b) Pixel-level neural-network classification separating mKate2 and mCherry into a color-coded map. 
(c) Confusion matrix for pixel classification (mKate2, mCherry, background), showing 96\% average accuracy. 
(d) Spectrally resolved regions of a single \emph{C.~elegans} tracked at sub-millisecond temporal resolution after fusion with DVS event data. Inset: worm schematic indicating mKate2 (green) and mCherry (red) labeling.}
\label{fig:spectral_image}
\end{figure}

Figure~\ref{fig:spectral_image} shows representative results. Diffractograms are classified into pixel-level spectral maps with 96\% average accuracy despite 23~nm emission-peak separation and >62\% spectral overlap. Spectral identity derived at the CMOS frame rate is assigned to DVS trajectories, enabling ultra-fast spectrally resolved tracking.

For live imaging, worms were placed in M9 buffer and recorded at an effective 100,000~frames/s using the DVS. Details of worm preparation and sample mounting are provided in the Supplement. Due to lower brightness of mKate2 and mCherry \cite{shaner_improved_2004,shcherbo_bright_2007}, 100~ms CIS exposure was used for high-SNR diffractograms. The DVS ($>$120~dB dynamic range) captured weak fluorescence transients with microsecond precision.

\subsection*{Blue-Light–Evoked Dynamics}

For tracking purposes, asynchronous event data are partitioned into discrete temporal windows, which serve as the input for a Hierarchical Agglomerative Clustering (HAC) algorithm. This stage groups events into distinct object candidates by evaluating their relative spatio-temporal proximity. These clustered centroids are subsequently integrated into a Kalman filter framework, which has been optimized via a systematic tuning process to maintain track continuity (refer to the Supplemental Material for details on the Optuna-based hyperparameter optimization). Although the filter assumes a constant velocity model, a high process noise covariance is implemented to robustly accommodate the non-linear and erratic locomotion characteristic of the biological samples. 

\begin{figure}[htb!]
\centering
\fbox{\includegraphics[width=0.95\linewidth]{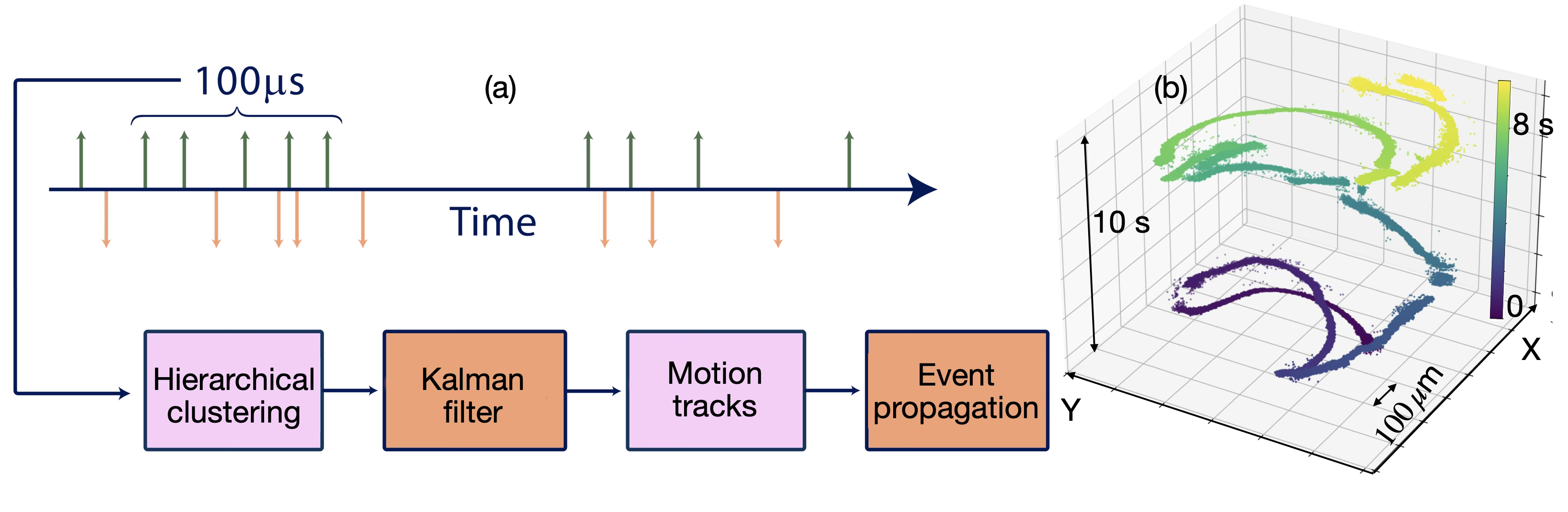}}
\caption{\textbf{Non-rigid motion tracking.}
(a) Overview of the event-based motion-tracking pipeline. Asynchronous DVS events are clustered, registered using Gaussian Mixture Model (GMM) point-set alignment, and temporally filtered with a Kalman estimator to produce continuous trajectories.  
(b) Representative tracking of a freely moving \emph{C.~elegans} within the field of view over $\sim$10~s.  }
\label{fig:tracking}
\end{figure}

Following the estimation phase, track identities are back-propagated to the constituent events, establishing a direct association between the raw signal and the tracked entities. This mapping enables the reconstruction of motion trajectories at arbitrary timescales, providing high-resolution insights into behavioral dynamics. This pipeline is illustrated in Fig.~\ref{fig:tracking}(a), while Fig.~\ref{fig:tracking}(b) demonstrates the pipeline's performance by tracking a fluorescently tagged elegan at a 100$\mu$s resolution across a continuous 12-second interval.

\begin{figure}[htb!]
\centering
\fbox{\includegraphics[width=0.95\linewidth]{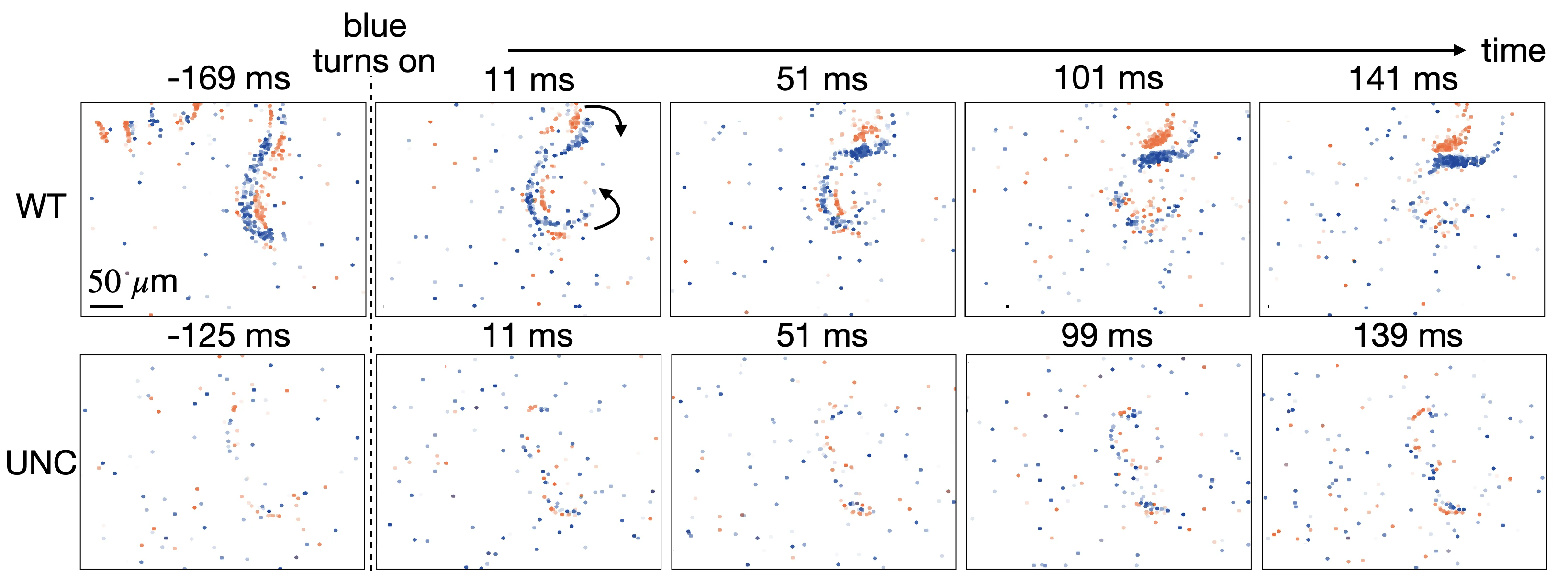}}
\caption{\textbf{Blue-light–evoked acceleration.}
Blue-light stimulation response of wild-type (WT) and mutant (UNC) worms in top and bottom panels, respectively. Events are accumulated over 10~ms temporal windows centered around the onset of blue excitation (dashed line indicates t = 0, when blue light turns on). The lower temporal resolution was chosen to match the fastest observed worm dynamics. Orange (positive-polarity) events mark pixels where intensity increased as the worm moved away, whereas blue (negative-polarity) events indicate intensity decreases due to occlusion as the worm moved toward those locations. When the two polarities remain spatially close (pre-stimulus), displacement within the accumulation window is minimal. Following blue-light onset, the separation between polarities increases, reflecting rapid acceleration and coiling behavior. Wild-type animals initiate this response within $\sim$0.18~s, whereas \emph{unc-101} mutants exhibit delayed coiling with latency approaching 0.88~s. Also see Supplementary videos 1 and 2.}
\label{fig:blue_effect}
\end{figure}

Worms were exposed to 460~nm step-function stimulation (500~ms off / 500~ms on). In DVS data, negative (blue) events indicate motion toward a pixel (occlusion), while positive (orange) events mark motion away. When polarities remain spatially close, displacement is small. Upon stimulus onset, the separation increases sharply, indicating acceleration in the blue-event direction (Fig.~\ref{fig:blue_effect}).

Wild-type (N2) animals exhibited canonical omega turns \cite{hori2018off,stirman2011real} in 7/12 trials, with mean latency $\sim$0.18~s. Non-responsive trials typically began mid-turn. In contrast, \emph{unc-101} mutants exhibited omega-like responses in 5/11 trials with delayed mean latency of 0.88~s (see exemplary videos in Supplementary Videos 1 and 2). Across trials, response latency differed between genotypes (two-sample test, $p<0.05$), although sample sizes were modest.

These results demonstrate that asynchronous–spectral fusion resolves stimulus-evoked acceleration and genotype-dependent behavioral differences with sub-millisecond temporal precision.

\section*{Conclusion}

We have introduced a fluorescence microscopy architecture that decouples spectral encoding from temporal sampling by fusing diffraction-enhanced CMOS imaging with asynchronous event-based sensing. The system achieves $\sim$23~nm spectral discrimination and effective temporal resolution of 100~$\mu$s while maintaining near-diffraction-limited spatial fidelity.

By assigning spectral identity directly to microsecond-resolved event trajectories, the proposed architecture enables spectrally resolved tracking at event-camera timescales without scanning or filter switching. In \emph{C.elegans}, this capability captures blue-light–evoked omega turns with high temporal fidelity and quantitatively resolves genotype-dependent response latencies (0.18s versus 0.88~s). Crucially, the asynchronous sensing paradigm provides a continuous, time-stamped data stream rather than discretized frames, establishing a strict upper bound on locomotion speed and eliminating ambiguity arising from inter-frame blind intervals. These results highlight the unique utility of event-based spectral imaging for interrogating rapid, behaviorally relevant dynamics that are otherwise inaccessible to conventional frame-based approaches.

Asynchronous–spectral fusion thus provides a scalable framework for ultrafast fluorescence imaging, extending temporal bandwidth into the sub-millisecond regime while preserving spectral fidelity. This capability opens new opportunities for interrogating rapid neural signaling, motor programs, and stimulus-evoked dynamics across biological systems.

\begin{backmatter}
\bmsection{Funding} The Chan Zuckerberg Initiative (CZI) grant: Dynamic-0000000282. E.M.J. is supported by the NIH grant NS034307 and is an investigator of the Howard Hughes Medical Institute. 

\bmsection{Acknowledgment} The authors thank Noor Syed for use of the EBIS camera and Lumos Imaging for the DOE. The support and resources from the Center for High Performance Computing at the University of Utah are gratefully acknowledged. Discussions with Al Ingold, Dajun Lin, Fernando Guevara-Vasquez and Fernando del Cueto are gratefully acknowledged.

\bmsection{Disclosures} The authors declare no conflicts of interest.

\bmsection{Data Availability Statement} Data and code underlying the results presented in this paper are available in https://github.com/Menonlab-Rich/hyperscope 

\bmsection{Supplemental document}
See Supplement 1 for supporting content.
\end{backmatter}

\bibliography{sample}

\bibliographyfullrefs{sample}

\end{document}